\documentclass[amsmath,amssymb,preprint,aps,nofootinbib]{revtex4}
\usepackage{natbib}
\def\beq{\begin{equation}}
\def\eeq{\end{equation}}
\def\bea{\begin{eqnarray}}
\def\eea{\end{eqnarray}}
\def\d{{\rm d}}

\begin{document}

\title{Superfluid Analog of the Davies-Unruh Effect}
\author{G.~L.~Comer}
\affiliation{Racah Institute of Physics, The Hebrew University, 
Jerusalem 91904, Israel}

\date{August, 1992}

\begin{abstract}
We produce an analog of the Davies-Unruh effect for superfluid helium 
four. \ There are two temperatures which result---one is associated with
ordinary, or first, sound and the other with second sound.
\end{abstract}

\maketitle

Two of the most striking results of the investigations into quantum field 
theory on curved backgrounds are the Hawking effect \cite{Hawking:1974sw} 
and the Davies-Unruh (D-U) effect 
\cite{Davies:1974th,unruh76:_notes}.\footnote{See also Fulling 
\cite{fulling72} for his preliminary contributions to the discovery of 
this effect.} \ The Hawking effect predicts that black holes radiate 
particles thermally. \ The D-U effect predicts that a system interacting 
with some quantum field and following a particular non-inertial path 
undergoes transitions as if it were immersed in a thermal bath of 
excitations of the given quantum field, even though the quantum field 
itself is in a global Minkowski vacuum state.

It is no coincidence that both results have thermal attributes. \ In each 
case that which is responsible for the thermal effects is an event 
horizon.\footnote{See Wald \cite{wald84:_gr} for a suitable description 
of event horizons, in particular, the importance of causality 
(information cannot be exchanged instantaneously between two spatially 
separated points) in their definitions.} \ In the black hole example the 
horizon is a feature of the geometric structure of the underlying 
spacetime; for the Davies-Unruh scenario a horizon exists because of the 
particular trajectory of the system. \ The main effect of a horizon is to 
delimit regions of spacetime between which information cannot be 
exchanged. \ In 1973, Bekenstein \cite{Bekenstein:1973ur} pointed out, 
for the particular example of black holes, that this inability to 
exchange information can bring about a violation of the second law of 
thermodynamics. \ He argued that this violation can be prevented if black 
holes have, in addition to their usual physical attributes, thermal 
characteristics (i.e.~entropy). \ The very general nature of Bekenstein's 
reasoning suggests that it should always be the case that additional 
thermal properties result whenever an event horizon is present.  

In 1981, Unruh \cite{Unruh:1980cg} generalized the understanding behind 
these effects when he proposed an analog to the Hawking effect for 
irrotational perfect fluids. \ He considered an ordinary fluid which is 
steadily accelerated from zero velocity to supersonic flow. \ The place 
at which the transition to supersonic flow occurs, the sonic horizon, is 
the fluid analog of an event horizon. \ For example, the region of the 
fluid beyond the sonic horizon cannot communicate---through the exchange 
of fluid excitations (phonons)---with the starting place of the fluid 
flow (i.e.~the fluid's ``asymptotic infinity'' where its velocity is 
zero).\footnote{Here, to ``communicate'' means to send information 
{\em only} through the exchange of phonons. \ Of course one could 
penetrate this horizon with something like a photon, say, but this should 
not ruin the event horizon nature of the sonic horizon for the problem at 
hand since it is interactions with the fluid excitations that dominate 
the end results.} \ Unruh found that the sonic horizon will emit phonons 
with a thermal distribution. \ The generalization occurs because here it 
is causality as defined by the paths that phonons take, not light, which 
underpins the physics.

An important technical aspect of Unruh's derivation is that one can 
associate the fluid notions of sonic horizon and asymptotic infinity with 
an underlying ``metric'' in the mathematics describing the system. \ Unruh
manipulates this metric in precisely the same way that real metrics are
manipulated (e.g., arbitrary coordinate transformations, construction of
metric compatible covariant derivatives, etc.) to obtain sonic horizon
thermal radiation. \ It is the main purpose of this work to apply Unruh's
technique in order to demonstrate an analog of the D-U effect for 
superfluid ${\rm He}^4$.

We use the two-fluid hydrodynamic model of superfluid ${\rm He}^4$ 
developed by Landau \cite{landau41}. \ The two fluids are the so-called 
normal- and super-fluids. \ The equations of motion for this model are 
(see Putterman \cite{putterman74:_book})
\bea
    0 &=& \rho_{,t} + \left(\rho_n v_{n j} + \rho_s v_{s j}\right)_{,j} 
          \ , \label{sfeq1} \\
    0 &=& \left(\rho s\right)_{,t} + \left(\rho s v_{n j}\right)_{,j} \ , 
          \label{sfeq2} \\
    0 &=& v_{si,t} + v_{sj} v_{si,j} + \mu_{,i} \ , \label{sfeq3} \\
    0 &=& \left(\rho_n v_{n i} + \rho_s v_{s i}\right)_{,t} + \left(p 
          \delta_{i j} + \rho_n v_{n i} v_{n j} + \rho_s v_{s i} v_{s j}
          \right)_{,j} \ , \label{sfeq4}
\eea
where the repeated indices are summed over 1, 2, and 3, a subscript comma
denotes partial differentiation, the subscripts n and s stand for normal-
and super-fluid variables, respectively, any $\rho$, with a subscript or
not, is a mass-density ($\rho$ without a subscript is the total 
mass-density), $v_j$ is the ${\rm j}^{\rm th}$ component of a 
three-velocity, $p$ is pressure, $s$ is the specific entropy, $\mu$ is 
the chemical potential, and $\rho_s \equiv \rho - \rho_n$. \ These 
variables are all functions of the Newtonian time $t$ and the spatial 
coordinates $x_j$. \ The eight equations of motion in 
Eqs.~(\ref{sfeq1})--(\ref{sfeq4}) are used to determine the eight 
independent degrees of freedom $\rho$, $s$, $v_{n i}$, and $v_{s i}$. \ 
To have a well-posed problem we must also supplement these equations with 
equations of state. \ Defining $w_i \equiv v_{n i} - v_{s i}$ and $w^2 
\equiv w_j w_j$, then the equations of state have the functional 
dependence \cite{putterman74:_book} $p = p[\rho,s,w^2]$, $\rho_n = 
\rho_n[\rho,s,w^2]$, and $\mu = \mu[\rho,s,w^2]$. \ If $T$ is the 
temperature, then the thermodynamic variables must also satisfy 
\cite{putterman74:_book}
\beq
     \d p = \rho \d \mu + \rho s \d T + \frac{1}{2} \rho_n \d w^2 \ . 
\eeq

The form of Eq.~(\ref{sfeq3}) presupposes that the superfluid flow is 
irrotational, i.e.~$\nabla \times \vec{v}_s = \vec{0}$. \ This implies 
that there exists a scalar potential $\phi_1$ such that
\beq
    v_{s i} = \phi_{1,i} \ . \label{irrot1}
\eeq
As a consequence of Eq.~(\ref{irrot1}), and an assumption about initial 
conditions, there is another vector field $A_i = (\rho_n w_i)/(\rho s)$ 
which is irrotational \cite{putterman74:_book}. \ Thus there is a scalar 
potential $\phi_2$ such that
\beq
    A_i = \phi_{2,i} \ . \label{irrot2}
\eeq

Because of the irrotational character of $v_{s i}$ and $A_i$ we really 
have only four independent degrees of freedom; therefore, some of the 
above equations are redundant. \ By inserting Eq.~(\ref{irrot1}) and 
Eq.~(\ref{irrot2}) into Eqs.~(\ref{sfeq1})--(\ref{sfeq4}) we can reduce 
the number of equations of motion from eight to four. \ Inserting 
Eq.~(\ref{irrot1}) into Eq.~(\ref{sfeq3}) yields
\beq
    0 = \left(\phi_{1,t} + \frac{1}{2} \phi_{1,j} \phi_{1,j} + \mu
        \right)_{,i} 
\eeq
so that
\beq
    0 = \phi_{1,t} + \frac{1}{2} \phi_{1,j} \phi_{1,j} + \mu + c_1(t) \ ,
        \label{phi1eq}
\eeq
where $c_1(t)$ is an arbitrary function of t. \ Likewise, we can reduce 
Eq.~(\ref{sfeq4}) to find
\beq
    0 = \phi_{2,t} + \frac{\rho s}{\rho_n} \phi_{2,j} \phi_{2,j} + 
        \phi_{1,j} \phi_{2,j} + T + c_2(t) \ , \label{phi2eq}
\eeq
where $c_2(t)$ is another arbitrary function of t. \ Now we have as our
independent degrees of freedom $\phi_1$, $\phi_2$, $\rho$, and $s$. \ 
Their equations of motion are Eqs.~(\ref{sfeq1}) and (\ref{sfeq2}) (with 
Eqs.~(\ref{irrot1}) and (\ref{irrot2}) inserted appropriately), 
Eq.~(\ref{phi1eq}), and Eq.~(\ref{phi2eq}), supplemented by the equations 
of state.

To demonstrate our analog of the D-U effect we use a linearized form of 
the equations of motion about some background solution.  \ The variables 
$\phi_1$, $\phi_2$, $\rho$, $s$, $p$, $T$, $\mu$, and $\rho_n$ all take 
the form $\tilde{\cal O} = {\cal O} + \delta {\cal O}$, where 
$\tilde{\cal O}$ represents a full solution, $\cal{O}$ represents the 
background, and $\delta {\cal O}$ is some variation away from the 
background. \ The prefactor $\delta$ stands for a quantity which is 
``small'' in the sense appropriate for this problem. \ We use a 
background which is homogeneous, isotropic, and static (i.e.~all 
background functions $\phi_1$, $\phi_2$, etc.~are independent of $t$ and 
$x_i$). \ This implies that the thermodynamic variables $p$, $T$, $\mu$, 
and $\rho_n$ are functions of only $\rho$ and $s$ since the background 
velocities vanish.

Before going any further it is convenient to relate some of the
thermodynamic derivatives which appear during the linearization with more
familiar quantities, i.e.~the heat capacities at constant volume $C_V = T 
\partial s/\partial T|_\rho$ and constant pressure $C_p = T \partial s/
\partial T|_p$  as well as the speeds of first and second sounds $u^2_1 = 
\partial p/\partial \rho|_s$ and $u^2_2 = (T s^2 \rho_s)/(C_V \rho_n)$, 
respectively \cite{putterman74:_book}. \ In terms of $C_V$, $C_p$, 
$u_1$, and $u_2$ we find
\bea
    \left.\frac{\partial p}{\partial s}\right|_\rho &=& - \frac{\rho}{s} 
          w^2_1 \ , \\
    \left.\frac{\partial \mu}{\partial \rho}\right|_s &=& 
          \frac{1}{\rho} \left[u^2_1 + w^2_1\right] \ , \\
    \left.\frac{\partial \mu}{\partial s}\right|_\rho &=& - \frac{1}{s} 
          \left[w^2_1 + \frac{\rho_n}{\rho_s} u^2_2\right] \ , 
\eea
where
\beq
    w^2_2 \equiv \left(\frac{\rho_s}{\rho_n} - 1\right) w^2_1 \equiv 
          \left(\frac{\rho_s}{\rho_n} - 1\right) u_1 u_2 \left(
          \frac{\rho_n}{\rho_s} \left[1 - \frac{C_V}{C_p}\right]
          \right)^{1/2} \ .
\eeq
Since the background velocities vanish we do not require any specific
formulas for the thermodynamic derivatives of $\rho_n$ (these derivatives 
will not appear in the linearized equations).

Inserting all variables which take the form $\tilde{\cal O} = {\cal O} + 
\delta {\cal O}$ into our equations of motion, keeping only terms linear 
in $\delta$, and using the various thermodynamic quantities identified 
above, we find
\bea
    \delta \rho &=& - \frac{\rho}{u^2_1} \left[1 - \frac{\rho_s}{\rho_n} 
                    \frac{w^4_1}{u^2_1 w^2_2}\right]^{- 1} \left[\delta 
                    \phi_{1,t} + s \left(1 + \frac{\rho_s}{\rho_n} 
                    \frac{w^2_1}{u^2_2}\right) \delta \phi_{2,t}\right] 
                    \ , \label{lineq1} \\
               && \cr
     \delta s &=& - \frac{s^2}{u^2_2} \frac{\rho_s}{\rho_n} \left[1 - 
                  \frac{\rho_s}{\rho_n} \frac{w^4_1}{u^2_1 u^2_2}
                  \right]^{- 1} \left[\left(1 + \frac{w^2_1}{u^2_1}
                  \right) \delta \phi_{2,t} + \frac{1}{s} 
                  \frac{w^2_1}{u^2_1} \delta \phi_{1,t}\right] \ , 
                  \label{lineq2} \\
        && \cr
     0 &=& - \frac{1}{u^2_1} \delta \phi_{1,tt} + \left[1 + 
           \frac{w^2_1}{u^2_1}\right] \delta \phi_{1,jj} + s \left[1 - 
           \frac{u^2_2}{u^2_1} - \frac{w^2_2}{u^2_1}\right] \delta 
           \phi_{2,jj} \ , \label{lineq3} \\
        && \cr
     0 &=& - \frac{1}{u^2_2} \delta \phi_{2,tt} + \left[1 - 
           \frac{w^2_1}{u^2_2}\right] \delta \phi_{2,jj} - \frac{1}{s} 
           \frac{w^2_1}{u^2_2} \delta \phi_{1,jj} \ . \label{lineq4} 
\eea
We have done some further manipulations to arrive at 
Eqs.~(\ref{lineq1})--(\ref{lineq4}): The linearized form of 
Eqs.~(\ref{phi1eq}) and (\ref{phi2eq}) did not involve any derivatives of 
$\delta \rho$ or $\delta s$ so they could be solved in terms of 
derivatives of $\delta \phi_1$ and $\delta \phi_2$ by purely algebraic 
means.

We now show that some of the terms which appear in 
Eqs.~(\ref{lineq1})--(\ref{lineq4}) can be neglected for temperatures $T 
= 10^{- 3}~{\rm K}$. \ In this temperature range the thermodynamics is 
dominated by phonons, the roton contributions being negligible 
\cite{london50:_superfl_supercond,london54:_superfl_He}, and $u_1 = 
\sqrt{3} u_2 = 2.4 \times 10^4~{\rm cm/sec}$. \ Using the various 
equations of states for superfluid ${\rm He}^4$ (see 
\cite{london50:_superfl_supercond,london54:_superfl_He}) and some basic 
thermodynamic identities, it can be shown that $s = 
u^2_1 \rho_n/T \approx 10^4~T^3~{\rm erg}/({\rm gm}~{\rm K}^4)$ 
and $(1 - C_V/C_p)^{1/2} \approx 10^{- 2}~T^2~{\rm K}^{- 2}$. \ We have 
assumed that $\rho_s$ is constant (i.e.~independent of $T$) so that 
$\partial \rho/\partial T|_p \approx \partial \rho_n/ \partial T|_p$. \ 
Using these approximations we find that the terms $w^2_1/u^2_1$, $w^2_1/
(s u^2_2)$, and $s [1 - (u^2_2 + w^2_2)/u^2_1]$ are all much less than 
one for $T < 10^{- 3}~{\rm K}$. \ Therefore, the equations for $\delta 
\phi_1$ and $\delta \phi_2$ decouple and take the following approximate 
form
\beq
    0 = - \frac{1}{u^2_a} \delta \phi_{a,tt} + \delta \phi_{a,jj} \ , 
        \label{waveq}
\eeq
where $a = 1~{\rm or}~2$ is just a label, not a tensor index.

Now we formally identify a metric from Eq.~(\ref{waveq}). \ This metric is
obviously that associated with a flat spacetime:
\beq
    \d s^2_a = - u^2_a \d t^2 + \d x_i \d x_i \ . \label{met}
\eeq
Note that time intervals are related to spatial intervals in this
``spacetime'' by the appropriate speed of sound $u_a$. \ The important 
structure here is the sonic-cone given by the congruence of curves which 
satisfy $\d s^2_a = 0$. \ It is this structure which allows for the 
existence of a horizon in superfluid ${\rm He}^4$ and underpins our 
Davies-Unruh analog. \ In contrast to real Minkowski spacetime, we see in 
this approximation to the dynamics of superfluid ${\rm He}^4$ two 
sonic cones---one for first sound and another for second 
sound.\footnote{We can anticipate similar results for third and fourth 
sound. \ Moreover, one can easily see that the Hawking effect analog will 
have more than one horizon (which will emit thermal excitations). 
\label{hefoot}} \ It is a general feature of 
Eqs.~(\ref{lineq1})--(\ref{lineq4}) that our system has two 
characteristic velocities for propagating excitations (it is not just a 
result of the approximations which generated Eq.~(\ref{waveq})) 
\cite{landau41,putterman74:_book,london50:_superfl_supercond,london54:_superfl_He}.

To extract our analog of the D-U effect we will use the arguments 
developed by Unruh in Ref.~\cite{unruh76:_notes}. \ They involve the 
construction of ``particle'' detection devices. \ These devices register 
excitations of the quantum field under consideration, for us $\delta 
\phi_a$. \ The detectors are to be accelerated, with constant proper 
acceleration $a_a$, along the z-axis. \ (In this context, ``proper'' 
quantities are formed with the ``metric'' in Eq.~(\ref{met}).) \ The 
worldlines traced out by the detectors have the following 
parametrization:
\beq
    z^2 - u^2_a t^2 = 4 R^2 \quad , \quad u_a t = z {\rm tanh}(\tau/2) 
      \ , \label{traj}
\eeq
and $x = y = 0$. \ The parameters $\tau$ and $R$ are the analogs of
Rindler coordinates for the metric in Eq.~(\ref{met}) \cite{rindler66}. \ 
In order to model how a detector registers excitations consider it to be 
a particle in a potential well. \ We will, for simplicity, let the 
potential be generated by a box. \ The box must confine the particle and 
at the same time permit it to interact with the fluid. \ Let the 
interaction of the particle with $\delta \phi_a$ be via the potential
\beq
    V_{int} = \varepsilon_a \delta \phi_a \Phi \ , 
\eeq
where $\Phi$ is the wave function of the particle and $\varepsilon_a$ is 
some coupling constant. \ Note that our choice of $V_{int}$ requires the
particle to interact either with $\delta \phi_1$ or $\delta \phi_2$ but 
not both.\footnote{In superfluid ${\rm He}^4$ it is possible to excite 
first sound modes without exciting second sound modes and vice versa. \ 
The reason for this is second sound modes are essentially thermal waves 
while first sound modes are associated with density disturbances. \ This 
follows from Eqs.~(\ref{lineq1}) and (\ref{lineq2}) since in the small 
$T$ limit $\delta \rho \propto \delta \phi_{1,t}$ and $\delta s \propto 
\delta \phi_{2,t}$. \ Also see Ref.~\cite{putterman74:_book}.} \ Also, 
since the gradient of $\delta \phi_a$ is a perturbation of a velocity, 
and velocities change sign under time reversal, $V_{int}$ is time 
antisymmetric. \ This is not a serious problem here since our main result 
will be seen to be independent of the particular form of the interaction 
potential (i.e.~no dependence on the coupling constant $\varepsilon_a$). 
\ The overall behavior of the particle is governed by its Schr\"odinger 
wave equation
\beq
    i \frac{\partial}{\partial \tau} \Phi = \frac{2 a_a}{u_a} 
    \frac{1}{2 m} \left(\frac{\partial^2}{\partial h^2} + 
    \frac{\partial^2}{\partial x^2} + \frac{\partial^2}{\partial y^2} + 
    m^2 \frac{u^2_a}{R_c} h + \varepsilon_a \delta \phi_a\right) \Phi \ ,
    \label{schro}
\eeq
where $m$ is the mass of the particle, $h \approx (R^2 - R^2_c)/R_c$ is 
its proper distance from the center $R_c$ of the box, $a_a = u^2_a/2 R_c$ 
is its proper acceleration, and the time derivative is with respect to the
proper time of the particle (hence, the reason for the extra factor $2 
a_a/u_a$). \ The term proportional to $m^2$ just represents the force 
experienced by the particle because it is accelerating.

Through some mechanism, which need not be discussed here, the detector is 
constrained to follow a $R = R_c = const$ trajectory in the $0 \leq z <
\infty$ and $- \infty < t < \infty$ region of a $u_a t - z$ spacetime 
diagram. \ As can be seen from Eq.~(\ref{traj}) this path is a hyperbola. 
\ The asymptotes of this hyperbolic path are generated by the sonic cone 
(i.e.~the $45^{\rm o}$ lines in our spacetime diagram) going through the 
origin ($t = z = 0$). \ It is not very difficult to see that the 
``future'' asymptote (i.e.~the one with positive slope) does represent an 
event horizon since a fluid excitation produced ``beyond'' it can never 
be registered by our detector.

The point of this construction is to see how the detector responds when 
it follows the given non-inertial path. \ A good indicator of this 
response is the transition probability per unit proper time $P_{a j}$ for 
the detector (i.e.~``particle'') to be excited from its ground state to 
some excited energy eigenstate with energy $E_j$. \ We will not present 
the details here, but rather we refer the reader to Unruh 
\cite{unruh76:_notes} since the calculation of $P_{a j}$ follows exactly 
as presented there.\footnote{The important steps are those which produce 
Eqs.~(3.1a) through (3.10) in Unruh \cite{unruh76:_notes}.} \ The final 
result is
\beq
    P_{a j} = \frac{1}{\exp\left(2 \pi u_a\left[E_j - E_0\right]/a_a
              \right) - 1} F_{det} \ , \label{trans}
\eeq
where $E_0$ is the ground state energy and $F_{det}$ just depends on the
characteristics of the detector ($\varepsilon_a$, etc.). \ Note that 
Eq.~(\ref{trans}) is precisely the form expected of a detector immersed 
in a thermal bath of excitations with temperature $T_a$ given by 
(restoring $\hbar$ and $k_b$)
\beq
    T_a = \frac{1}{2 \pi} \frac{\hbar}{k_b} \frac{a_a}{u_a} \ . 
          \label{temp}
\eeq
This is the desired result. \ (As claimed earlier, it is independent of 
$\varepsilon_a$.) \ For each speed of sound $u_a$ there is an associated 
temperature (cf.~footnote \ref{hefoot}). \ Notice that this is the same 
formula obtained by Davies and Unruh except that the speed of sound $u_a$ 
replaces the speed of light. \ It is, up to factors of order one, what 
one would obtain from a dimensional analysis, using only the relevant 
physical parameters of this problem. \ Putting in the values for $\hbar$, 
$k_b$, and $u_a$ gives $T_a \approx 10^{- 13}~(a_a/g)~{\rm K}$ where $g 
\approx 10^3~{\rm cm}/{\rm sec}^2$.

That we obtain two different temperatures is a necessary consequence of 
the fact that there are two different sonic-cone structures in superfluid 
${\rm He}^4$. \ This is a required consistency check of the principle 
underlying our D-U analog---two sonic-cone structures necessarily implies 
two types of event horizons, which, in turn, suggests two different 
thermal effects.

To arrive at Eq.~(\ref{temp}) we have formally identified certain 
characteristics of superfluid ${\rm He}^4$ with an underlying metric 
present in the mathematics. \ We then formally used this metric in ways 
that are familiar from differential geometry and quantum field theory on 
curved backgrounds. \ At best, our derivation can only be considered 
heuristic since it is not obvious how to interpret the ``proper time'' 
$\tau u_a/2 a_a$ appearing in Eq.~(\ref{schro}). \ It becomes especially 
problematic if one imagines a non-relativistic ion being the ``detector'' 
described by Eq.~(\ref{schro}). \ The accepted understanding is that it 
is only Newtonian time, like $t$ in Eqs.~(\ref{sfeq1})--(\ref{sfeq4}), 
which is the ion's proper time. \ If $\tau u_a/2 a_a$ has any real 
meaning it must be statistical, in essence, and perhaps, as Unruh states 
for his Hawking effect analog, ``... (have) no physical significance 
except for the small-amplitude sonic motions of the medium'' 
\cite{Unruh:1980cg}. \ Whatever the interpretation, if one accepts the 
premise that horizons and thermal effects necessarily occur together, 
then one has little recourse but to believe that $T_a$ must have real, 
and not merely formal, significance. \ The problem, then, is to figure 
out how it can be detected.

I would like to thank J. D. Bekenstein for introducing me to Unruh's 
irrotational fluid calculations and for suggesting that I look at 
superfluids. \ His guidance and insight were crucial for the completion of
this work. \ I also wish to thank T.~Piran, C.~A.~Fuchs and M.~Schiffer 
for their support and Amos Ori for useful discussions in which some of the
conceptual difficulties of this work were clarified. \ This research was
supported by a Golda Meir fellowship with the Hebrew University of
Jerusalem, Israel.

\bibliography{biblio}

\section*{Appendix: Note Added in 2005}

The work presented here is unpublished, and not otherwise available. \ 
The original preprint is of some historic interest: it was one of the 
earliest articles in the modern, post-1990 revival of ``analog models'' 
for curved spacetime. \ Every attempt has been made to maintain the 
original form of the preprint. \ One typographical error, in 
Eq.~(\ref{sfeq4}), has been fixed. \ The referencing and footnoting are 
also different, since a modern typesetting program has been used.

The author wishes to thank Matt Visser and Haret Rosu for citing this 
work on a few occasions, and thus preventing it from passing away 
like Lennon and McCartney's Eleanor Rigby. \ He also thanks several 
colleagues and collaborators for encouraging him to post it on the 
e-print archive.

The author's current address is Department of Physics, Saint Louis 
University, St.~Louis, MO 63156-0907, USA.

\end{document}